\documentclass[conference]{IEEEtran}
\IEEEoverridecommandlockouts

\usepackage{cite}
\usepackage{amsmath,amssymb,amsfonts}
\usepackage[noend]{algorithmic}
\usepackage{algorithm}
\usepackage{graphicx}
\usepackage{textcomp}
\usepackage{xcolor}
\usepackage{hyperref}
\usepackage{orcidlink}
\usepackage{enumitem}
\usepackage{booktabs}
\usepackage{amsmath}
\usepackage{array}
\usepackage{subcaption}
\newcolumntype{Y}{>{\centering\arraybackslash}m{1.6em}}

\newcommand{\name}{\textsc{Raven}\xspace}
\newcommand{\rb}{\textsc{RavenBERT}\xspace}

\usepackage{xcolor}

\usepackage{multirow}

\usepackage{balance}
\usepackage{siunitx}
\usepackage{expl3,xparse} 
\usepackage[most]{tcolorbox}
\tcbuselibrary{listings} 
\usepackage{cleveref}
\usepackage{soul} 
\usepackage{listings} 
\usepackage{hyperref}
\usepackage{xspace}

\newcommand{\code}[1]{{\scriptsize\texttt{#1}}}

\usepackage{float}

\usepackage{tabularx}
\usepackage{caption}

\lstdefinestyle{mystyle}{
    basicstyle=\small\ttfamily,
    breakatwhitespace=false,         
    breaklines=true,                 
    postbreak=\mbox{\textcolor{red}{$\hookrightarrow$}\space},
}
\makeatletter
\newcommand{\TODO}[1]{%
 \bgroup
 \def\@tempa{#1}%
 \expandafter\textcolor\expandafter{red}{\@tempa}%
 \GenericWarning{}{LaTeX Warning: TODO: \@tempa}%
 \egroup
}
\makeatother

\makeatletter
\newcommand{\NOTE}[1]{%
 \bgroup
 \def\@tempa{#1}%
 \expandafter\textcolor\expandafter{blue}{\@tempa}%
 \GenericWarning{}{LaTeX Warning: NOTE: \@tempa}%
 \egroup
}
\makeatother

\usepackage{soul}

\usepackage{tikz}
\usetikzlibrary{arrows.meta,patterns,calc}
\usepackage{tabularx}
\usepackage{tcolorbox}

\tcbset{
  pefttile/.style={
    enhanced,
    colback=white, colframe=black,
    rounded corners,
    boxsep=1.5pt, left=3pt, right=3pt, top=3pt, bottom=3pt,
    title style={font=\scriptsize\bfseries},
  }
}

\usepackage{enumitem}
\setlist{nosep,leftmargin=*,itemsep=0pt,topsep=2pt,parsep=0pt,partopsep=0pt}

\usepackage[font=small]{caption}
\definecolor{lightgreen}{rgb}{0.85, 1.0, 0.85}
\definecolor{lightred}{rgb}{1.0, 0.85, 0.85}

\newcounter{finding}
\crefname{finding}{Finding}{Findings} 
\Crefname{finding}{Finding}{Findings} 

\tcbset{
  flamesfinding/.style={
  breakable,
    colback=blue!3!white,
    colframe=blue!50!black,
    coltitle=white,
    fonttitle=\bfseries,
    colbacktitle=blue!70!black,
    enhanced,
    sharp corners=all,
    attach boxed title to top center={yshift=0mm},
    boxed title style={
        size=small,
        colframe=blue!50!black,
        colback=blue!70!black,
        sharp corners=all,
    },
    drop shadow southeast,
    boxrule=0.2mm,
    arc=0.5mm,
    top=0.7mm,
    bottom=0.7mm,
    left=0.7mm,
    right=0.7mm,
    boxsep=0.5mm,
  }
}

\usepackage{listings, xcolor}

\definecolor{verylightgray}{rgb}{.97,.97,.97}

\lstdefinelanguage{Solidity}{
	keywords=[1]{anonymous, assembly, assert, balance, break, call, callcode, case, catch, class, constant, continue, constructor, contract, debugger, default, delegatecall, delete, do, else, emit, event, experimental, export, external, finally, for, function, gas, if, implements, import, in, indexed, instanceof, interface, internal, is, length, library, log0, log1, log2, log3, log4, memory, modifier, new, payable, pragma, private, protected, public, pure, push, require, return, returns, revert, selfdestruct, send, solidity, storage, struct, suicide, super, switch, then, this, throw, transfer, try, typeof, using, value, view, while, with, addmod, ecrecover, keccak256, mulmod, ripemd160, sha256, sha3}, 
	keywordstyle=[1]\color{blue}\bfseries,
	keywords=[2]{address, bool, byte, bytes, bytes1, bytes2, bytes3, bytes4, bytes5, bytes6, bytes7, bytes8, bytes9, bytes10, bytes11, bytes12, bytes13, bytes14, bytes15, bytes16, bytes17, bytes18, bytes19, bytes20, bytes21, bytes22, bytes23, bytes24, bytes25, bytes26, bytes27, bytes28, bytes29, bytes30, bytes31, bytes32, enum, int, int8, int16, int24, int32, int40, int48, int56, int64, int72, int80, int88, int96, int104, int112, int120, int128, int136, int144, int152, int160, int168, int176, int184, int192, int200, int208, int216, int224, int232, int240, int248, int256, mapping, string, uint, uint8, uint16, uint24, uint32, uint40, uint48, uint56, uint64, uint72, uint80, uint88, uint96, uint104, uint112, uint120, uint128, uint136, uint144, uint152, uint160, uint168, uint176, uint184, uint192, uint200, uint208, uint216, uint224, uint232, uint240, uint248, uint256, var, void, ether, finney, szabo, wei, days, hours, minutes, seconds, weeks, years},	
	keywordstyle=[2]\color{teal}\bfseries,
	keywords=[3]{block, blockhash, coinbase, difficulty, gaslimit, number, timestamp, msg, data, gas, sender, sig, value, now, tx, gasprice, origin},	
	keywordstyle=[3]\color{violet}\bfseries,
	keywords=[4]{true},
	keywordstyle=[4]\color{green!50!black},
	keywords=[5]{false},
	keywordstyle=[5]\color{red!80!black},
	identifierstyle=\color{black},
	sensitive=false,
	comment=[l]{//},
	morecomment=[s]{/*}{*/},
	commentstyle=\color{gray}\ttfamily,
	stringstyle=\color{red}\ttfamily,
	morestring=[b]',
	morestring=[b]"
}

\definecolor{precolor}{HTML}{4169E1}    
\definecolor{vulncolor}{HTML}{FF4500}   
\definecolor{postcolor}{HTML}{228B22}   

\colorlet{precolorlight}{precolor!25}
\colorlet{vulncolorlight}{vulncolor!25}
\colorlet{postcolorlight}{postcolor!25}

\newcommand{\conditionDotBox}[3]{%
  \begingroup
  \renewcommand{\arraystretch}{0}%
  \setlength{\tabcolsep}{0pt}%
  \begin{tabular}{@{}c@{}}%
    \colorbox{precolorlight}{\makebox[0.4em][c]{\ifnum#1=1\textcolor{black}{$\bullet$}\else\phantom{$\bullet$}\fi}} \\%
    \colorbox{postcolorlight}{\makebox[0.4em][c]{\ifnum#3=1\textcolor{black}{$\bullet$}\else\phantom{$\bullet$}\fi}}%
  \end{tabular}%
  \endgroup
}

\usepackage{array,booktabs}
\newlength{\TripColW}
\newlength{\ModelSep}
\newlength{\ModelBlockW}
\newcolumntype{B}{>{\centering\arraybackslash}m{\ModelBlockW}}

\newcommand{\flames}{\textsc{Flames}\xspace}

\newcommand{\sindi}{\textsc{Sindi}\xspace}

\DeclareRobustCommand{\code}[1]{\lstinline[basicstyle=\ttfamily\small]{#1}}

\usepackage{inconsolata}
\DeclareRobustCommand{\code}[1]{\lstinline[basicstyle=\ttfamily\small]{#1}}

\DeclareRobustCommand{\tablecode}[1]{\lstinline[basicstyle=\ttfamily]{#1}}

\def\BibTeX{{\rm B\kern-.05em{\sc i\kern-.025em b}\kern-.08em
    T\kern-.1667em\lower.7ex\hbox{E}\kern-.125emX}}
\begin{document}









\title{\name: Mining Defensive Patterns in Ethereum via Semantic Transaction Revert Invariants Categories}

\author{
\IEEEauthorblockN{
Mojtaba Eshghie\orcidlink{0000-0002-0069-0588}\textsuperscript{1},
Melissa Mazura\textsuperscript{2},
Alexandre Bartel\textsuperscript{1},
}
\IEEEauthorblockA{\textsuperscript{1}Umeå University, Sweden}
\IEEEauthorblockA{\textsuperscript{2}\textit{KTH Royal Institute of Technology}, Stockholm, Sweden}
\thanks{Emails: \{mojtabae, alexandre.bartel\}@cs.umu.se; mazura@kth.se.}
}


\maketitle

\begin{abstract}
We frame Ethereum transactions reverted by an invariant--\code{require(<invariant>) / assert(<invariant>) / if (<invariant>)} \code{revert} statements in the contract implementation--as a positive signal of active on-chain defenses. Despite their value, the defensive patterns in these
transactions remain undiscovered and underutilized in security research. We present \name, a framework that aligns reverted transactions to the invariant causing the reversion in the smart
contract source code, embeds these invariants using our BERT-based fine-tuned model, and clusters them by semantic intent to mine defensive invariant categories on Ethereum. Evaluated on a sample of \num{20000} reverted transactions, \name achieves cohesive and meaningful clusters of transaction-reverting invariants. Manual expert review of the mined 19 semantic clusters uncovers six new invariant categories absent from existing invariant catalogs, including feature toggles, replay prevention, proof/signature verification, counters, caller-provided slippage thresholds, and allow/ban/bot lists. To demonstrate the practical utility of this invariant catalog mining pipeline, we conduct a case study using one of the newly discovered invariant categories as a fuzzing oracle to detect vulnerabilities in a real-world attack. \name thus can map Ethereum’s successful defenses. These invariant categories enable security researchers to develop analysis tools based on data-driven security oracles extracted from the smart contract’s working defenses.
\end{abstract}


\section{Introduction}

Smart contract invariants are used as oracles in verification tools~\cite{ItyFuzz,OracleSupportedExploitGen,xplogen,dynamit}, runtime monitors~\cite{highguard}, or deployable defenses within the contract's implementation~\cite{flames}. 
They are implemented using \code{require(<invariant>)} / \code{assert(<invariant>)} / \code{if (<invariant>)} \code{revert/throw;} statements in Solidity language~\cite{EthereumYellowPaper2022,solidity_docs}, and can revert the atomic execution of a transaction on the blockchain. They enforce security properties such as access control, value bounds for asset transfers, and temporal constraints. Literature identifies the existence of millions of unique invariant implementations in Solidity contracts on the Ethereum~\cite{DISL}. Yet our understanding of which defensive patterns in these invariants actually work in the wild remains limited.

Despite billions of dollars lost to incidents and attacks~\cite{SoK}, research on decentralized application security mostly involves post-mortem analyses of incidents (defense mechanism failures). Meanwhile, the blockchain continuously emits a richer signal: \emph{reverted} transactions. Every reverted transaction due to an invariant is a concrete instance where a contract's defense was triggered—often reflecting a prevented exploit, misuse, bot behavior, or economically unprofitable trade~\cite{Solana}. Up to November 2nd, 2025, Ethereum contains \num{71379564} failed transactions to smart contracts (out of the total more than 3 billion transactions recorded on-chain)~\cite{totalEthFailedContractTxs}. However, it is not clear which invariants or invariant categories caused these reverts. This creates a paradox where the literature studies successful exploits~\cite{SoK,zhou_sok_2023} and replicates them~\cite{xplogen,AIAgentExploitGen,OracleSupportedExploitGen}, but the defenses that successfully prevented the attacks and malicious interactions are understudied.

Research on smart-contract invariants has advanced in dynamic invariant mining from successful transactions~\cite{DemystifyingInvariants,SmartOracle,InvCon,InvCon+}, static invariant synthesis~\cite{flames,InvSol,VeriSmart}, and vulnerability detection using invariants~\cite{highguard,dynamit,ItyFuzz}. These approaches rely on expert-curated invariant catalogs as templates or training data~\cite{TrustLLM,flames,InvCon+,InvCon,DemystifyingInvariants}, yet none systematically use real-world transaction-reverting invariants as actionable defenses. 
Existing reverted-transaction studies examine only surface-level patterns such as emitted events~\cite{Solana}, while Liu et al.~\cite{CharacterizingTXRevertingSC} provide a code-level analysis of \num{3866} contracts, but do not link their static findings to real on-chain reversion reasons. Thus, no comprehensive categorization connects reverted transactions to the source-level invariants that triggered them.

We present \name\footnote{\url{https://github.com/mojtaba-eshghie/Raven/}}
, a tool to identify semantic clusters of transaction revert invariants. \name (1) collects revert invariants from the source code of Ethereum contracts with failed transactions 
(2) neurally embeds them, and (3) clusters the embedded invariants by {intent} (e.g., access control, value bounds, etc.).
Organizing reversion invariants into coherent categories, \name provides an empirically grounded view of the Ethereum ecosystem's {working defenses}.

This paper makes the following contributions:
\begin{itemize}[leftmargin=*,nosep]
\item \rb: A BERT-based embedding model\footnote{\url{https://huggingface.co/MojtabaEshghie/RavenBERT}}
fine-tuned on a set of revert-inducing transaction invariants on Ethereum extracted from \num{100}k failed transactions. 
\item A methodology to discover the best configuration for semantic clustering of invariants.
\item A catalog of \num{19} categories of invariants found by \name, and labeled by manual review, resulting in the discovery of \emph{six} previously unreported defensive invariant categories. 
\item A dataset of 20k failed Ethereum transactions attributes, such as the reversion invariants and their localization information in the source code\footnote{\url{https://huggingface.co/datasets/MojtabaEshghie/raven-dataset}}, such as reversion invariant, function, and contract.

\item A case study demonstrating the practical utility of \name's findings by using the newly discovered ``proof verification'' invariant category to construct a fuzzing oracle that successfully detects the Nomad Bridge incident's vulnerability.

\end{itemize}


Our study is guided by three research questions: 
\begin{itemize}[leftmargin=*,nosep]
  \item \textbf{RQ1 (Clustering Intrinsic Quality Evaluation).} Does \name generate compact, well-separated groups of invariants based on well-known intrinsic clustering quality metrics? \emph{Answer:} \name's best configurations produce compact, well-separated clusters of invariants with a Silhouette score of $0.93$ and S\_Dbw score of $0.043$.
  \item \textbf{RQ2 (Coverage–Quality Trade-off).} What fraction of invariants is clustered (non-noise), and how does this trade off with the semantic quality of clustering? \emph{Answer:} \name shows a clear coverage–quality trade-off with the highest semantic clustering quality at $52$–$58\%$ coverage of invariants. 
  
  \item \textbf{RQ3 (Meaningfulness of Mined Invariant Groups).} Does \name generate meaningful invariant groupings based on human reviews? \emph{Answer:} Manual analysis reveals that the clusters are meaningful and non-overlapping. Partial overlap of generated cluster labels with previous invariant catalogs shows the clustering validity, and the newly discovered \emph{six} invariant categories show the semantic effectiveness of both \name's design and parameter grid search methodology.

  \item \textbf{RQ4 (Effectiveness in Vulnerability Detection).} Can the invariant categories discovered by \name be used to detect vulnerabilities in real-world contracts? \emph{Answer:} A case study on the Nomad Bridge incident demonstrates that invariant categories discovered by \name (specifically, proof verification checks) can be effectively translated into actionable fuzzing oracles to detect complex upgrade vulnerabilities.
\end{itemize}

Our work presents the first study on the feasibility of using clustering on transaction revert invariants to understand the defensive patterns behind Ethereum blockchain contracts.

The rest of the paper is organized as follows. \Cref{sec:arch} presents \name's design. \Cref{sec:datasets} provides the datasets used in our experiments. \Cref{sec:evaluation_protocol} details the evaluation protocol for \name.  \Cref{sec:results} presents the results while \Cref{sec:discussion} discusses the implication of the results, threats to validity, and open questions.
\Cref{sec:related_works} reviews related works and \Cref{sec:conclusion} concludes the paper.

\section{System Architecture}\label{sec:arch}

\begin{figure}[t]
    \centering
    \includegraphics[width=0.9\linewidth]{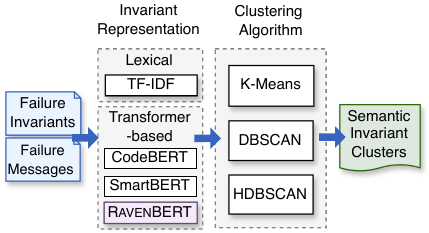}
    \caption{\name's pipeline for semantic clustering of revert-inducing transaction invariants.}
    \label{fig:raven_arch}
\end{figure}

Grouping invariants causing the transactions to revert on Ethereum provides insights into underlying contract defenses.
To achieve this, we design the semantic clustering pipeline of \name in \Cref{fig:raven_arch} that uses both lexical 
and neural embedding models, including \rb for invariant feature representation (see \Cref{sec:inv_representation}). After embedding the invariants, we employ clustering algorithms to discover semantic transaction-reverting invariant groups (\Cref{sec:Clustering}).

\subsection{Invariant Representations}\label{sec:inv_representation}
To develop a comprehensive embedding space, we consider two feature representations: i) a lexical token-based embedding, TF-IDF~\cite{TF_IDF}, which counts the occurrence of tokens in the input sequence, and ii) encoder-only transformer models based on BERT~\cite{bert}, which transform input into neural representations.
When embedded in these spaces, semantic similarity between data points corresponds to geometric proximity~\cite{ClusteringTechniques}, and enables clustering based on similarity metrics. 
We use cosine similarity to capture semantic differences between invariants, as it has been successfully utilized in prior works~\cite{SolBERT}.
Using both token-based vectorization and transformer-based embeddings enables a rigorous comparison of their impact on the quality of clustering.

\subsubsection{Lexical Vectorization}  
We utilize the Term Frequency-Inverse Document Frequency (TF-IDF)~\cite{TF_IDF} vectorization to transform Solidity/Vyper invariants into sparse vector representations. Due to its capability to detect Type 1 and Type 2 clones, which do not require deep code understanding, this embedding serves as a string-based baseline for other representations to compare against~\cite{Siamese}.

\subsubsection{Transformer-Based Embedding}
To capture richer semantic and syntactic features of invariants, we investigate three pretrained transformer models designed for code understanding~\cite{AttentionIsAllUNeed}. 
Out of the encoder, decoder, and encoder-decoder transformer model families, we only use the encoder family for embedding models. 
All three of our tested encoder-only models belong to the BERT family~\cite{bert}. 

\noindent\textbf{CodeBERT:} A general code model based on the BERT transformer used for a baseline of the embedding models.

\noindent{\textbf{SmartBERT-v2}}: This is a domain-specific model for smart contract sources~\cite{smartbert2024}. SmartBERT-v2 is pre-trained on a corpus of \num{16}k Solidity files.

\noindent{\textbf{\rb}}: Pretrained models have proven effective for capturing semantic representations, but they lack invariant-specific domain knowledge. Therefore, we use SmartBERT-v2 as a smart contract-aware pre-trained model and further fine-tune it on failure-inducing invariants. Other models, such as SolBERT~\cite{SolBERT} and SmartEMBED~\cite{SmartEmbed}, are also tailored to Solidity code through CFG/AST representations of the program, which makes them unsuitable for ad hoc invariant fine-tuning (they are not trained directly on source code). 
Motivated by the fine-tuning approaches employed in related models, such as CodeBERT~\cite{CodeBERT}, which utilizes Masked Language Modeling (MLM), and SolBERT~\cite{SolBERT}, which leverages contrastive learning, we experimented with both methods to identify the most effective fine-tuning approach for \name. 
We first employed {Masked Language Modeling (MLM)}~\cite{bert}, a pretraining technique where random tokens in the input sequence are masked, and the model learns to predict them using bidirectional context. MLM is the standard pretraining objective for BERT-family models and has proven effective for general-purpose code understanding tasks. We hypothesized that further MLM fine-tuning on invariant-specific code could help the model capture domain-specific token distributions and syntactic patterns unique to guard predicates. However, despite training with reduced epochs (down from 3 to 1) to prevent overfitting, MLM fine-tuning resulted in slightly worse performance compared to the base SmartBERT-v2 model. This suggests that MLM's token-level prediction objective, while effective for pretraining, does not directly optimize for the semantic similarity structure needed for clustering tasks on invariants.
Next, we employed {contrastive learning}~\cite{ContrastiveLearning}, a self-supervised technique that learns representations by contrasting similar (positive) and dissimilar (negative) pairs of data points. The core principle is to pull semantically similar items closer together in the embedding space while pushing dissimilar items apart, thereby creating a representation space where cosine similarity reflects semantic relatedness. 
To implement contrastive learning for \name, we generated pseudo-labeled pairs by computing cosine similarity across our dataset of invariants. We used a similarity threshold (\num{0.8}) to automatically create positive pairs (semantically similar invariants) and negative pairs (semantically dissimilar invariants). We generated \num{16470} contrastive pairs of invariants using this approach. These artificially labeled pairs served as ground truths during the fine-tuning process, guiding the model to learn invariant-specific semantic distinctions beyond what SmartBERT-v2's contract-level pretraining provided. This approach proved to be significantly better in terms of embedding quality and is the foundation of \name's success in representing smart contract invariants. We perform L2 normalization after embedding the predicates to ensure all vectors lie on the unit sphere, making cosine distance a proper metric for clustering.

\subsection{Clustering Algorithm}\label{sec:Clustering}
To systematically group invariants and detect previously undiscovered groups of invariants, we employ a suite of clustering algorithms to determine the most effective one. 
Semantic clustering is a technique for data analysis used in Natural Language Processing (NLP)~\cite{Sem_Clustering}, which groups phrases together based on their semantic similarity.
Clustering has evolved into distinct types. Partitioning clustering~\cite{ClusteringTechniques} divides data points into \(k\) clusters based on an objective function. K-Means~\cite{KMeans} is one such widely used algorithm that divides data by minimizing the within-cluster variance. 
Another widely used algorithm is Density-Based Spatial Clustering of Applications with Noise (DBSCAN)~\cite{DBSCAN} which requires~\num{2} parameters: the neighborhood radius \(\varepsilon\) and the minimum number of points to form a cluster. Density-based clustering partitions data based on regions of high density, using a threshold~\cite{ClusteringTechniques}.
Moreover, hierarchical DBSCAN (HDBSCAN~\cite{ran2023comprehensive}) clustering algorithm forms a tree-like structure to capture data groups by iteratively splitting data based on similarity.  
K-Means~\cite{KMeans} is chosen for its effectiveness with spherical, well-separated clusters, and DBSCAN~\cite{DBSCAN} is utilized for its performance for clustering with noise. HDBSCAN~\cite{HDBSCAN} is an improvement over DBSCAN through hierarchical analysis, providing improved performance in scenarios with varying cluster densities. 
\name therefore uses these three clustering algorithms: K-Means~\cite{KMeans}, DBSCAN~\cite{DBSCAN}, and HDBSCAN~\cite{HDBSCAN}.


\begin{figure}[t]
    \centering
    \includegraphics[width=0.9\linewidth]{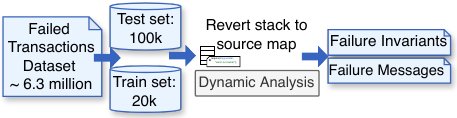}
    \caption{Failure invariant dataset collection pipeline.}
    \label{fig:dataset_collection}
\end{figure}

\section{Dataset Collection}\label{sec:datasets}
To train and evaluate \name, we collected two datasets. We use the first for fine-tuning our embedding model, \rb, and the second dataset is used for the end-to-end evaluation of \name. Both of these are subsets of a total of~\num{6313121} failed transactions on Ethereum in the time span of June~15th, 2024 (block~\num{20100158}), until March~28th, 2025 (block~\num{22145000}).
To prevent data leakage from training to test, we ensured non-overlapping training and test sets.

\noindent\textbf{Fine-tuning Dataset:} To provide a high number of invariants for fine-tuning the model, this collection includes \num{100000} randomly selected failed transactions. Utilizing the invariant extraction pipeline in \Cref{fig:dataset_collection} to accumulate the data yielded~\num{1932} unique invariants (using string similarity).

\noindent\textbf{Evaluation Dataset:} Given the computational demand of the dynamic analysis necessary for retrieving transaction invariants at scale, instead of fine-tuning/evaluating \name on millions of failed Ethereum transactions to date, we employ a sampling strategy to strike a balance between feasibility based on resources and statistical significance.
To determine a statistically representative sample size, we employed Cochran's formula~\cite{StatisticShowToCochran,CochranPaper}. We defined our parameters for a~\num{95}~\,\% confidence level and a ±~\num{1}~\,\% margin of error. We remained conservative by assuming maximum population variability (p =~\num{0.5}), which is the standard practice when the true distribution of the data is unknown. The calculation yielded a required minimum sample of~\num{9604} transactions from the population of~\num{6313121}. The actual sample used in our evaluation is~\num{20000} transactions, which substantially surpasses the mentioned minimum requirement.

\begin{figure}
    \centering
    \includegraphics[width=0.9\linewidth]{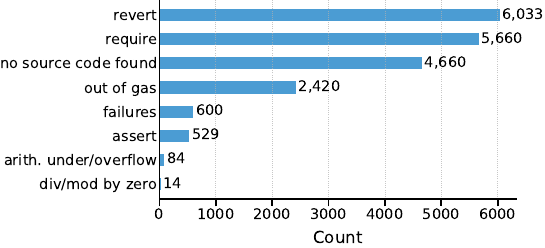}
    \caption{Summary of the failure categories in the test set.}
    \label{fig:failure_types}
\end{figure}

\subsection{Transaction Invariant Extraction}
The dataset collection begins by retrieving basic transaction information for each transaction hash. This includes the transaction's sender address, receiver address, gas used, gas price, gas limit, timestamp, block number, value, transaction index, and transaction input. This stage also gives us a preliminary transaction failure reason, which helps us filter the ones that will not constitute a transaction reverted with an invariant, ruling out cases that do not require further dynamic analysis. These preliminary errors fall into two categories: 1) ``Out of gas'' failures, where the transaction exceeded its allowed gas limit. These have the \code{"Out of gas"} string in their error messages. 2) Arithmetic failures: These include overflow, underflow, and division by zero. Since Solidity 0.8.0, these are automatically reverted upon detection~\cite{solidity0.8Changes}. 

If the transaction does not fall into one of the above, we run a dynamic analysis tool to extract the exact stack frame that reverts the transaction and retrieve the invariant in the source code based on it, using bytecode-to-source mapping. We conduct the dynamic analysis step using Tenderly's APIs~\cite{tenderlyapi}. The mined information in this step includes the line number and contract identifier where the error originated. For verified contracts, we map line numbers to the source code file and validate the correctness of the determined line number and invariant by Tenderly based on the fetched verified source code from Etherscan. 
Therefore, we skip and remove the failed transactions where the ``to'' address is not a verified contract. We flag and report these transactions with ``no source code''.
When tracing the reverted transaction's last stack frame resolves to a valid LoC in the verified source code, we extract the predicate from one of the three statements in Solidity/Vyper sources: \code{assert}, \code{require}, or \code{if (...) revert/throw}, with the addition of the optional error message in case of \code{require/assert}.
 

After extracting the invariants, we de-duplicate their set to have a unique dataset of failed transaction invariants. 
The de-duplication process reduced the sample to~\num{727} unique invariant patterns (\num{3.6}~\%). Among these,~\num{396} 
contained error messages. Invariant strings with error messages averaged~\num{53.9} characters in length, while those without averaged~\num{37.4} characters.

\subsection{Failure Invariants Characteristics}
\label{sec:dataset_characteristics}

As \Cref{fig:failure_types} shows, out of \num{20}k initial randomly selected failed transactions, we extracted~\num{12222} invariants (\num{61.1}\,\%) (revert, require, and assert bars in \Cref{fig:failure_types}). Extraction failures 
comprising~\num{600} instances (\num{3}\%). All other (non-invariant) transaction reversion reasons contained~\num{7178} cases (\num{35.89}\%). Among non-invariant failures, ``No source code found'' is the largest category with~\num{4660} entries (\num{23.3}\%), followed by ``Out of gas'' errors at~\num{2420} items (\num{12.1}\%).
Authors manually validated a randomly-selected subset of \num{1000} transactions to ensure the extraction results are valid.

\begin{table}[t]
\centering
\scriptsize
\renewcommand{\arraystretch}{1}
\caption{Schema of the collected dataset.}
\begin{tabularx}{\linewidth}{@{} l X @{}}
\toprule
\textbf{Dataset Attribute} & \textbf{Description} \\
\midrule
\texttt{hash} & Transaction hash \\
\texttt{failure\_reason} & Short description of why the transaction failed \\
\texttt{block\_number} & Block number of the transaction \\
\texttt{from\_address} & Sender's Ethereum address \\
\texttt{to\_address} & Receiver's Ethereum address \\
\texttt{tx\_input} & Input data sent with the transaction \\
\texttt{gas} & Gas used by the transaction \\
\texttt{gas\_price} & Gas price specified for the transaction \\
\texttt{gas\_limit} & Maximum gas allowed \\
\texttt{value} & Ether value transferred \\
\texttt{tx\_index} & Transaction's position within its block \\
\texttt{failure\_message} & Error message from simulation \\
\texttt{failure\_invariant} & Extracted condition that caused the failure \\
\texttt{failure\_file} & File path or source location of the failure \\
\texttt{failure\_function} & Name of function where failure occurred \\
\texttt{failure\_contract} & Contract address involved in failure \\
\texttt{timestamp} & Timestamp of the transaction \\
\bottomrule
\end{tabularx}
\label{tab:dataset_schema}
\end{table}


\subsection{Final Dataset Contents}
\Cref{tab:dataset_schema} presents the final dataset schema containing detailed information about failed Ethereum transactions, including metadata (e.g., timestamp and transaction index), participants (sender and receiver), execution context (gas, input, value), and failure attributes (reason, message, invariant, source availability, and revert localization details in the source code).

\section{Evaluation Protocol}\label{sec:evaluation_protocol}

This section defines how we search for the best configuration of \name to generate meaningful and reliable clusters.
After fine-tuning \rb, we use a grid search to find the optimal clustering parameters of \name. We then generate the semantic clusters for the test dataset and compute the metrics defined in \Cref{sec:metrics} to answer RQ1--RQ2. \Cref{sec:rq3_protocol} presents the protocol for manually analyzing the identified semantic clusters (RQ3), and \Cref{sec:rq4_protocol} outlines our case study protocol (RQ4).   

\subsection{Experimental Setup}\label{sec:experimental_setup}

We use the non-overlapping sets from \S\ref{sec:datasets}:
(i) a \emph{fine-tuning} set of \num{100000} failed transactions yielding \num{1932} unique invariants (used only to train \rb), and
(ii) an \emph{evaluation} sample of \num{20000} failed transactions, from which we extract and deduplicate predicates into \num{727} unique invariants.
We consider two \emph{views} for each extracted invariant: \textit{predicate-only} and \textit{predicate+message} (when an error string exists).
In the embedding layer, neural models use their native tokenizers, and embedding vectors are L2-normalized. 

After running the experiments, we compare the four encoders defined in \S\ref{sec:datasets}–\S\ref{sec:Clustering}: TF–IDF, CodeBERT, SmartBERT-v2, and \rb (contrastively fine-tuned on both invariant views). 
We use the cosine distance as the metric for semantic similarity in clustering. During the clustering, density-based clustering algorithms (DBSCAN and HDBSCAN) assign a label of $-1$ to noise points.

\subsection{Evaluation Metrics (RQ1 and RQ2 Protocols)}\label{sec:metrics}
A ground-truth labeling of invariants does not exist (prior dynamic miners like Trace2Inv~\cite{DemystifyingInvariants} and InvCon+~\cite{InvSol,InvCon+} report invariant categories but do not provide comprehensive taxonomies of smart contract invariants). Hence, external metrics like ARI~\cite{ARI}, which tells how often the clustering and the ground truth agree, require labels that are unavailable. We therefore resolve to use four robust \emph{intrinsic} metrics of the generated clusters jointly to select the best-performing configuration of \name without access to ground truths: Silhouette, S\_Dbw, Coverage, and Admissibility.

\subsubsection{Silhouette~\cite{SilhouetteDefinition} (bounded \([-1,1]\), higher is better)} For point \(i\),
\begin{equation}
Sil(i)=\frac{b(i)-a(i)}{\max\{a(i),b(i)\}}
\end{equation}

where \(a(i)\) is the mean intra-cluster distance and \(b(i)\) the minimum mean distance to any other cluster. 

\subsubsection{S\_Dbw~\cite{sdbw}, lower is better)}\label{subsec:sdbw}
A single score that prefers clusters in which (i) points inside a cluster sit close together in embedding space (intra-cluster tightness) and (ii) clearly separated (few points lying between two clusters).

\begin{equation}
\resizebox{0.98\columnwidth}{!}{%
$ \displaystyle
\mathrm{S\_Dbw}
=
\underbrace{\frac{1}{k}\sum_{i=1}^{k}
\frac{\mathrm{scatter}(C_i)}{\mathrm{scatter}(\mathcal{D})}}_{\text{within-cluster tightness}}
+
\underbrace{\frac{1}{k(k-1)}\sum_{i\ne j}
\frac{\mathrm{dens}(\mathrm{mid}(c_i,c_j))}
{\max\{\mathrm{dens}(c_i),\,\mathrm{dens}(c_j)\}}}_{\text{between-cluster overlap}}
$%
}
\end{equation}


where \(\mathrm{scatter}(C_i)\) is the average \emph{cosine} distance of points in cluster \(i\) to its centroid \(c_i\); \(\mathrm{dens}(p)\) counts points within a fixed radius \(r\) around \(p\); we set \(r\) to the mean within-cluster scatter across all clusters. \(\mathrm{mid}(c_i,c_j)\) is the midpoint on the unit sphere between centroids \(c_i\) and \(c_j\).
We compute this using cosine distances on L2-normalized embeddings. For DBSCAN/HDBSCAN, we exclude noise points and count only non-empty clusters in \(k\). Compared to the Silhouette score, which is intuitive but can be optimistic for roughly spherical partitions, S\_Dbw further penalizes loose clusters and overlap between neighboring clusters; therefore, using both provides a more balanced judgment.

\begin{table}[t]
    \centering    
    \caption{Hyperparameter ranges for grid search.}
    \scriptsize
    \begin{tabular}{c|c}
        \toprule
        Algorithm & Parameters \\
        \midrule
        K-Means  & $k \in [8, 100]$ \\
        DBSCAN   & $\epsilon \in [0.1, 5]$, $min\_samples \in [10, 15]$\\
        HDBSCAN  & $\epsilon \in [0.1, 1]$, $min\_cluster\_size \in [10, 15]$ \\
        \bottomrule
    \end{tabular}
    \label{tab:clustering_params}
\end{table}

\subsubsection{Coverage}
Coverage captures the extent to which a clustering algorithm confidently structures the space. Density methods (e.g., DBSCAN/HDBSCAN) can trade coverage for cluster homogeneity by labeling uncertain points (e.g., border points between clusters) as noise, thereby decreasing coverage. 

\begin{equation}
\text{Coverage}= \frac{\#\left\{\substack{\text{invariants assigned} \\ \text{to a non-noise cluster}}\right\}}{\#\{\text{unique invariants}\}} \times 100\%.
\end{equation}

\subsubsection{Admissibility and Best Model Selection}
We define a configuration triple admissible if sufficies all the three following criteria:
\begin{itemize}[leftmargin=*,nosep]
  \item \emph{C1 (Granularity):} \(8 \le \#\text{clusters} \le 100\) (motivated below),
  \item \emph{C2 (Coverage floor):} Coverage \(\ge 50\%\) over invariants,
  \item \emph{C3 (Stability):} metrics unchanged to two decimals over two re-runs (same seed) to guard against nondeterminism.
\end{itemize}


\noindent\textbf{Hyperparameter Search.} 
For each configuration triple (invariant embedding, clustering algorithm, invariant view), we grid-search to find admissible settings through the hyperparameter ranges defined in \Cref{tab:clustering_params}. 
The \({k\geq8}\) lower bound (\Cref{tab:clustering_params}) follows the eight top-level invariant categories observed by Trace2Inv~\cite{DemystifyingInvariants}; the \({k\leq100}\) cap preserves human inspectability. 
The cluster number \emph{k} corresponds to the minimum and maximum number of clusters.
Among admissible runs (C1--C3) for a fixed configuration triple, we select one configuration by priority: first minimize S\_Dbw; if tied to two decimals, maximize Silhouette; if still tied, maximize Coverage; if still tied, prefer fewer clusters. This yields a single best-per-triple setting that balances cohesion/separation, assignment rate, and parsimony of clustering.
Revert strings used in \code{require/assert} statements (e.g., OpenZeppelin's Ownable contract~\cite{openzeppelin-access-2x}: \code{require(msg.sender==owner,"caller is not the owner")}) 
recur across many contract libraries and can dominate clustering by surface form rather than predicate semantics. We therefore evaluate both \emph{predicate-only} and \emph{predicate+message} views to detect and mitigate template-driven clustering artifacts during our grid search for the best hyperparameters.


\subsection{Protocol for Manual Analysis (RQ3)} \label{sec:rq3_protocol}

To evaluate the semantic meaningfulness of the invariant groups generated by our best configuration found via RQ1 and RQ2 protocol, we conduct a manual expert review of the generated clusters.  
We conduct a qualitative analysis involving two authors. In case of a conflict of opinions between the first two authors, the third author's vote determines the verdict.
The review process consists of three phases:

\subsubsection{Semantic Labeling} For every cluster generated by the best configuration, the annotators independently inspect the invariants to identify the common intent of the invariants in the cluster (e.g., checking token allowance vs. checking time constraints). A cluster is deemed \textit{meaningful} if a clear, unifying semantic logic applies to the majority (\>\num{90}\%) of its constituent invariants. The authors then assign a descriptive category label to the cluster. 

\subsubsection{Taxonomy Alignment and Novelty Detection} To determine if \name discovers defensive patterns overlooked by prior work, we map our derived labels against the invariant templates and taxonomies from three state-of-the-art invariant mining and synthesis tools: Trace2Inv~\cite{DemystifyingInvariants} (23 templates), InvCon+~\cite{InvCon+} (14 templates), and the formalized business logic properties from DCR graphs~\cite{FormalizingDCRMojtaba}. A cluster is classified as a \textit{new category} if its semantic intent does not map to any existing template in these catalogs.

\subsubsection{Cluster Distinction} Finally, we verify that the generated clusters are distinct from one another. We analyze the semantic boundaries between clusters to ensure that separate clusters represent non-overlapping logic (e.g., invariants with intents ``Caller Not Owner'' and ``Contract Paused'' are not merged).

\subsection{Protocol for Real-World Incident Analysis (RQ4)}\label{sec:rq4_protocol}


To demonstrate that a newly discovered invariant category by \name can act as a {practical oracle} in detecting a real-world vulnerability, we conduct a case study. This case study is conducted based on the RQ3 results categories.
We begin by scanning incident reports in DeFiHackLabs~\cite{DeFiHackLabs} to identify one incident with characteristics suggesting that it could have been prevented if one of our newly discovered invariant categories (as presented in RQ3 results) had been used to analyze the contract.
We use the matching category to construct a fuzzing oracle for this specific incident.
Based on this oracle, we create a Foundry-based fuzz test that randomly creates transaction sequences targeting the upgrade procedure of the identified incident. Our success in detecting the incident's  
vulnerability using this fuzz test 
demonstrates the real-world effectiveness of the discovered category in preventing the incident.

\section{Results}\label{sec:results}


We report clustering results on \num{20}k failed transactions (June~2024--March~2025), comparing two invariant views, four embeddings, three algorithms.
\Cref{sec:clustering_performance,sec:best_config_analysis} present quantitative and manual analysis of \name. \Cref{sec:rw_incident} presents the results of our case study.

\subsection{Clustering Performance}\label{sec:clustering_performance}
Running the hyperparameter grid search detailed in \Cref{sec:evaluation_protocol} results in the {best admissible} results for the \num{24} configurations of \name presented in \Cref{tab:clustering-results-error,tab:clustering-results-no-error}.

\subsubsection{Silhouette Score Analysis}

DBSCAN achieved the highest silhouette scores (\num{0.79}--\num{0.93}), with \name\ reaching \num{0.93} (no messages) and \num{0.89} (with messages). HDBSCAN ranged \num{0.59}--\num{0.85} (SmartBERT best at \num{0.85}), while K-Means scored \num{0.49}--\num{0.68} (CodeBERT worst at \num{0.49}).

\begin{table}[t]
\centering
\scriptsize
\renewcommand{\arraystretch}{1.1}
\caption{Clustering evaluation results with error messages.}
\label{tab:clustering-results-error}
\begin{tabularx}{\linewidth}{@{} p{1.2cm} p{1.05cm} p{0.25cm} p{0.5cm} p{0.05cm} p{0.4cm} >{\raggedright\arraybackslash}X @{}}
\hline
\textbf{Embedding} & \textbf{Alg.} & \textbf{Sil.} & \textbf{S\_Dbw} & \textbf{\#C} & \textbf{\%C} & \textbf{Search-Mined Params} \\[0.2mm]
\hline
TF-IDF                & K-Means   & 0.55 & 0.034 & 99 & 100.0 &  n\_clusters: 99 \\
{TF-IDF}       & {DBSCAN}   & {0.88} & {0.063} & {17} & {56.67} &  {eps:0.36,min\_samples:10} \\
{TF-IDF}       & {HDBSCAN}  & {0.76} & {0.067} & {11} & {81.71} &  {min\_cluster\_size:13} \\
CodeBERT              & K-Means   & 0.52 & 0.061 & 99 & 100.0 &  n\_clusters: 99 \\
CodeBERT              & DBSCAN   & 0.83 & 0.079 & 18 & 56.53 &  eps:0.36, min\_samples:10 \\
CodeBERT              & HDBSCAN  & 0.59 & 0.115 & 11 & 76.62 &  min\_cluster\_size:12 \\
SmartBERT             & K-Means   & 0.57 & 0.071 & 99 & 100.0 &  n\_clusters:99 \\
SmartBERT             & DBSCAN   & 0.84 & 0.090 & 16 & 54.8 &  eps:0.36,min\_samples:12 \\
SmartBERT             & HDBSCAN  & 0.65 & 0.141 & 16 & 72.08 &  min\_cluster\_size:10 \\
\rb                & K-Means   & 0.60 & 0.028 & 99 & 100.0 &  n\_clusters:99 \\
\textcolor{blue}{\textbf{\rb}}       & \textcolor{blue}{\textbf{DBSCAN}}   & \textcolor{blue}{\textbf{0.89}} & \textcolor{blue}{\textbf{0.042}} & \textcolor{blue}{\textbf{18}} & \textcolor{blue}{\textbf{58.32}} &  \textcolor{blue}{\textbf{eps:0.36,min\_samples:14}} \\
\rb                & HDBSCAN  & 0.73 & 0.058 & 21 & 83.77 &  min\_cluster\_size: 10 \\
\bottomrule
\end{tabularx}
\end{table}

\subsubsection{S\_dbw Score Analysis}

\rb with DBSCAN clustering algorithm achieved best S\_dbw scores (\num{0.042} with messages, \num{0.043} without), outperforming HDBSCAN (\num{0.056}--\num{0.141}) and K-Means (\num{0.028}--\num{0.078}). Manual inspection revealed K-Means over-divides semantic groups into \num{99} clusters (max allowed). Therefore, we exclude it from further analysis.

\begin{figure*}[t]
    \centering
    \scriptsize
    \begin{subfigure}[b]{0.27\linewidth}
        \centering
        \includegraphics[width=\linewidth]{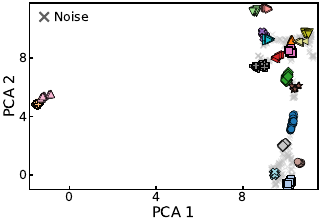}
        \caption{\name}
        \label{fig:rebert_without_pca}
    \end{subfigure}
    \hfill 
    \begin{subfigure}[b]{0.28\linewidth} 
        \centering
        \includegraphics[width=\linewidth]{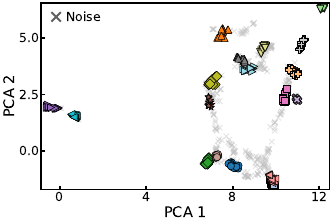}
        \caption{\name\ + msgs.}
        \label{fig:rebert_with_pca}
    \end{subfigure}
    \hfill 
    \begin{subfigure}[b]{0.28\linewidth}
        \centering
        \includegraphics[width=\linewidth]{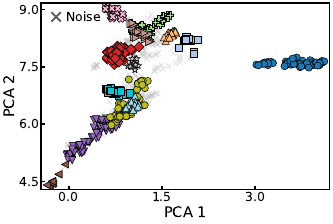}
        \caption{SmartBERT}
        \label{fig:smartbert_without_pca}
    \end{subfigure}

    \caption{PCA Analysis for the clusters generated by the best-performing configurations of \name.}
    \label{fig:combined_clusters} 
\end{figure*}

\subsubsection{Coverage Analysis}
DBSCAN covered \num{51.7}--\num{58.3}\% (14--20 clusters), with \name\ achieving highest at \num{58.3}\%. HDBSCAN reached \num{61.3}--\num{85.9}\% (10--23 clusters). 

\subsubsection{Optimal Clustering Strategies}
Performance analysis yielded three superior configurations based on combined S\_dbw score, silhouette score, and coverage (bold rows in \Cref{tab:clustering-results-error} and \Cref{tab:clustering-results-no-error}).
\rb without error messages using DBSCAN achieves a Silhouette score of~\num{0.93}, a coverage of~\num{51.86}\%, and an S\_dbw score of~\num{0.043}. 
\name\ with messages using DBSCAN achieves a Silhouette score of~\num{0.89}, a coverage of~\num{58.32}\%, and an S\_dbw score of~\num{0.042}. 
SmartBERT without messages with DBSCAN achieves a Sil. score of~\num{0.86}, a coverage of~\num{54.6}\%, and a S\_dbw score of~\num{0.069} and yields the next-best performance after \rb.

\refstepcounter{finding}
\begin{tcolorbox}[title=RQ1 Answer: Intrinsic Quality Evaluation, flamesfinding, label=finding:intrinsic_quality]
\name's best configuration found via our grid search, is \rb without error messages using DBSCAN algorithm with Silhouette score \num{0.93} and S\_Dbw score \num{0.42}.
\end{tcolorbox}

\begin{table}[t]
\centering
\scriptsize
\renewcommand{\arraystretch}{1.1}
\caption{Clustering results without error messages.}
\label{tab:clustering-results-no-error}
\begin{tabularx}{\linewidth}{@{} p{1.15cm} p{1.05cm} p{0.25cm} p{0.5cm} p{0.05cm} p{0.38cm} >{\raggedright\arraybackslash}X @{}}
\hline
\textbf{Embedding} & \textbf{Alg.} & \textbf{Sil.} & \textbf{S\_Dbw} & \textbf{\#C} & \textbf{\%C} & \textbf{Search-Mined Params} \\[0.2mm]
\hline
TF-IDF                & K-Means   & 0.55 & 0.078 & 99 & 100.0   & n\_clusters: 99 \\
{TF-IDF}       & {DBSCAN}   & {0.90} & {0.095} & {14} & {51.72}   & {eps:0.36,min\_samples:10} \\
TF-IDF                & HDBSCAN  & 0.80 & 0.136 & 10 & 61.35   & min\_cluster\_size: 17 \\
CodeBERT              & K-Means   & 0.49 & 0.049 & 99 & 100.0   & n\_clusters: 99 \\
CodeBERT              & DBSCAN   & 0.79 & 0.065 & 18 & 55.16   & eps:0.36, min\_samples:11 \\
CodeBERT              & HDBSCAN  & 0.73 & 0.064 & 13 & 62.04   & min\_cluster\_size: 12 \\
SmartBERT             & K-Means   & 0.52 & 0.054 & 99 & 100.0   & n\_clusters: 99 \\
SmartBERT             & DBSCAN   & 0.86 & 0.069 & 20 & 54.61   & eps:0.36,min\_samples:11 \\
\textcolor{blue}{\textbf{SmartBERT}}    & \textcolor{blue}{\textbf{HDBSCAN}}  & \textcolor{blue}{\textbf{0.85}} & \textcolor{blue}{\textbf{0.062}} & \textcolor{blue}{\textbf{12}} & \textcolor{blue}{\textbf{77.99}}   & \textcolor{blue}{\textbf{min\_cluster\_size:10}} \\
\rb                & K-Means   & 0.68 & 0.030 & 99 & 100.0   & n\_clusters:99 \\
\textcolor{blue}{\textbf{\rb}}       & \textcolor{blue}{\textbf{DBSCAN}}   & \textcolor{blue}{\textbf{0.93}} & \textcolor{blue}{\textbf{0.043}} & \textcolor{blue}{\textbf{19}} & \textcolor{blue}{\textbf{51.86}}   & \textcolor{blue}{\textbf{eps:0.36,min\_samples:15}} \\
\rb\                & HDBSCAN  & 0.76 & 0.056 & 23 & 85.97   & min\_cluster\_size:10 \\
\bottomrule
\end{tabularx}
\end{table}

\subsection{Analysis of \name's three Best Clustering Configurations}\label{sec:best_config_analysis}
Here we analyze the produced clusters of the \emph{three} superior model configurations of \name. 

\subsubsection{PCA Analysis}\label{sec:pca_analysis}
Principal Component Analysis (PCA) is a linear projection that rotates the embedding space to align with axes (principal components), capturing the maximal variance. Plotting the first two components provides a 2D sketch of the geometry in high dimensions, revealing how tight clusters are, how far apart they are, and where ambiguous border points and noise reside. In our unsupervised setting with no ground truth, PCA serves as a sanity check and interpretation aid that complements intrinsic metrics (Silhouette, S\_Dbw). Compact and well-separated islands for each cluster in the PCA plot correspond to high Silhouette (high cohesion, high separation), which in turn makes the clustering semantically valid.
As \Cref{fig:rebert_without_pca} shows, \textsc{Raven}'s embedding+DBSCAN with predicate-only view clusters form small, dense islands with clear gaps between them, while a sizeable set is left as noise. This matches the strong intrinsic quality metric results (Silhouette \num{0.93}, S\_Dbw \num{0.043}) at conservative coverage ($\approx\num{52}\%$, \Cref{tab:clustering-results-no-error}). This is useful for reliable taxonomy-building tasks (tight, unambiguous semantic groups). Clusters of \textsc{RavenBERT}+DBSCAN with error messages in \Cref{fig:rebert_with_pca} visibly stretch more along PC1. Here, coverage rises
, but Silhouette is lower compared to its counterpart without error messages. 
This suggests error messages inject additional variance (often stylistic/template-like), creating more border cases. If messages added crisp semantic distinctions, we would expect higher Silhouette. Lastly, in SmartBERT+HDBSCAN with predicate-only view in \Cref{fig:smartbert_without_pca}, more points are absorbed into clusters leading to higher coverage ($\approx\num{78}$\%) but the PCA view shows elongated clusters, softer boundaries, and more possible overlaps compared to the previous two configurations, consistent with its intrinsic metric results (Silhouette \num{0.85}). These suggest that compact, well-separated clusters allow us to make strong, defensible claims about what each cluster represents and align with \name's goal of discovering meaningful semantic categories, rather than covering all possible cases (i.e., the cluster's semantic confidence over higher coverage). 
Furthermore, the added variance introduced by the messages in \rb embedding and DBSCAN configuration means they boost coverage but slightly degrade cluster cohesion by incorporating lexical information. This allows us to conclude that the triple (\rb embedding, DBSCAN, predicate-only view) is the best configuration. Next, we conduct a manual expert analysis of all of the generated clusters of this configuration (\Cref{sec:manual_review}).

\refstepcounter{finding}
\begin{tcolorbox}[title=RQ2 Answer: Coverage–Quality Trade-off, flamesfinding, label=finding:tradeoff]
We observe a clear trade-off between coverage and quality. \name yields the highest quality at $52$–$58\%$ coverage while some configurations raise the coverage to $72$–$86\%$ with lower cluster quality (grouping semantically unrelated invariants). 
Based on the trade-off,
 the configuration triple (\name's own model, DBSCAN clustering algorithm, and predicate-only invariant views) is selected as the best way to semantically cluster invariants. 
 Our results show that including the failure message increases coverage (\num{52}\% $\rightarrow$ \num{58}\%) at the expense of less reliable semantic clustering.
\end{tcolorbox}

\begin{table*}[t]
\centering
\scriptsize
\renewcommand{\arraystretch}{1.1}
\setlength{\tabcolsep}{4pt}
\caption{Cluster labels (\rb embedding, DBSCAN, predicate-only) with one representative invariant per cluster. Rows 2, 6, 7, 9, 12, and 14 are newly discovered categories that are absent from the literature.}
\label{tab:raven_dbscan_clusters}
\begin{tabularx}{\linewidth}{@{} c p{2.70cm} X  p{6.3cm} c @{}} 
\toprule
\textbf{\#} & \textbf{Assigned category} & \textbf{Example(s) from the cluster} & \textbf{Description} & \textbf{\# of TXs} \\
\midrule
1  & Wallet budget guards & \tablecode{balanceof(from) >= amount} & Ensure sufficient balance, allowance, and reserves before ops. & (7.40\%) \num{515} \\

\textcolor{blue}{\textbf{2}}  & Slip-safe thresholds & \tablecode{amountout >= amountoutmin} & Slippage guards ensure received amounts meet min thresholds. & (44.78\%) \num{3116} \\

3  & Tax/fee bypass privileges & \tablecode{isexcludedfromfees[from]||isexcludedfromfees[to]} & 
OR-based exemptions from transaction fees and taxes. & (7.86\%) \num{547} \\

4  & Time constraints & \tablecode{block.timestamp > deadline} & Deadline and expiration validations using block timestamps. & (6.47\%) \num{450} \\

5  & Non-zero sanity & \tablecode{reservein > 0 \&\& reserveout > 0}& -  & (4.51\%) \num{314} \\

\textcolor{blue}{\textbf{6}}  & Feature toggles & \tablecode{tradeenabled}, \tablecode{!launched}, \tablecode{!data.tradingenabled} & Positive boolean flags requiring enabled/active features. & (4.28\%) \num{298}  \\

\textcolor{blue}{\textbf{7}}  & Replay prevention & \tablecode{!usedclaims[claimleaf]} & Ensuring actions are not repeated (claims, mints, submissions). & (0.62\%) \num{43} \\

8  & Not-paused/not-stopped & \tablecode{!paused} & Inverted boolean checks for paused/disabled/inactive states. & (2.27\%) \num{158} \\

\textcolor{blue}{\textbf{9}}  & Proof/signature verification & \tablecode{!merkleproof.verify(proof, merkleroot, leaf)} & Merkle proof, signature, and hash verification operations. & (0.75\%) \num{52} \\

10  & Sender budget guards & \tablecode{balances[msg.sender] >= _amount} & Message sender's balance and spending approval checks. & (0.80\%) \num{56} \\

11 & Payment sanity checks & \tablecode{_amount * price == msg.value} & -  & (0.92\%) \num{64} \\

\textcolor{blue}{\textbf{12}} & (Allow/ban/bot/white)-list & \tablecode{!isblacklisted[msg.sender]} & Include/exclude lists controlling who can interact. & (1.24\%) \num{86} \\

13 & Cooldowns/Rate Limits & \tablecode{cooldown[to] < block.timestamp} & Block-/time-based throttling between operations. & (0.83\%) \num{58} \\

\textcolor{blue}{\textbf{14}} & Counters/nonces & \tablecode{allowed.nonce != nonce} & Nonce tracking and state sequence number enforcement. & (1.62\%) \num{113} \\

15 & Mint/supply caps & \tablecode{totalsupply() + mint_amount <= max_supply} & Maximum supply ceiling enforcement for token creation. & (0.60\%) \num{42} \\

16 & Access control & \tablecode{owner() == _msgsender()} & The most basic forms of access control. & (1.18\%) \num{82} \\

17 & Address sanity & \tablecode{recipient != address(0)} & Non-zero addresses, not \tablecode{this} contract, valid pair endpoints. & (1.06\%) \num{74} \\

18 & Low-level input validations & \tablecode{success\&\&(data.length==0||abi.decode(data,(bool)))} & Multi-clause low-level input validation logic. & (3.48\%) \num{242} \\

19 & Budget floors & \tablecode{balance >= amount} & Straightforward budget floors. & (9.31\%) \num{648} \\
\hline
& {Total clustered transactions} & & & \num{6958} \\
\bottomrule
\end{tabularx}
\end{table*}

\subsubsection{Manual Review of the Generated Clusters}
\label{sec:manual_review}
We manually inspected all clusters produced by the best configuration (\rb, DBSCAN, predicate-only view), and assigned semantic labels (see \Cref{tab:raven_dbscan_clusters}). Our review confirms that most clusters fall into well-known families (e.g., Access Control, Data/Data-Flow, Temporal/Time). It also reveals \emph{six} invariant categories that are \emph{absent from prior invariant mining tools' templates}---specifically, from Trace2Inv's enumerated \num{23}-template catalog, InvCon+'s \num{14} invariant templates, and the \num{15} formalized business logic properties modeled by DCR graphs~\cite{DemystifyingInvariants,InvCon+,FormalizingDCRMojtaba}. Below we list these categories:

\begin{itemize}[leftmargin=*, nosep, label=\checkmark, labelsep=3pt]  
\item \emph{Caller-provided slippage thresholds (relational)} (Cluster~2): guards like \code{amountOut >= amountOutMin} that relate two \emph{runtime} variables as slippage guard. Trace2Inv's~\cite{DemystifyingInvariants} ``oracle slippage'' uses constant bounds; InvCon+ and DCR graphs formalizations do not contain a relatable category~\cite{InvCon+,captureDCR}. 
  \item \emph{Feature toggles} (Clusters~6): contract-wide enable/disable toggles such as \code{tradeEnabled}. These are not covered by Trace2Inv's~\cite{DemystifyingInvariants} Access Control (EOA/owner/manager) nor by InvCon+'s~\cite{InvCon+}, and do not match the ``pausable'' pattern formalizations~\cite{FormalizingDCRMojtaba}.
  \item \emph{Replay prevention} (Cluster~7): one-time/consume-once checks (e.g., \tablecode{!usedClaims[leaf]}) that forbid re-use of claims/mints; no counterpart in prior catalogs~\cite{InvCon,InvCon+,DemystifyingInvariants,captureDCR}.
  \item \emph{Proof/signature verification} (Cluster~9): Merkle and signature validation (\code{MerkleProof.verify}) absent in prior works.
  \item \emph{Counters/nonces} (Cluster~14): explicit sequence enforcement which is not present in prior works.
  \item \emph{Allow/ban/bot/whitelist gates} (Cluster~12): list-membership gating (e.g., \code{!isBlacklisted[msg.sender]}), conceptually distinct from owner/manager checks in the prior works.
\end{itemize}

\noindent{Combined, these newly discovered clusters account for \num{3708} transactions reversions in our test set, representing \num{30.33}\% of the test set transactions that were reverted by an invariant (\num{3708}/\num{12222}).}

\refstepcounter{finding}
\begin{tcolorbox}[title=RQ3 Answer: Meaningfulness of Mined Patterns, flamesfinding, label=finding:meaningfulness]
Expert analysis of the generated invariants by \name's best configuration reveals: 1) consistency and overlap of a portion of the resulting clusters with previous invariant taxonomies and catalogs~\cite{DemystifyingLoops,InvCon+,captureDCR} and 2) identifying \emph{six} meaningful invariant categories ($\approx\num{30}\%$ of invariant-based reverted transactions in test set) previously ignored. Furthermore, expert analyses conclude that all generated clusters are meaningful and distinct from one another.
\end{tcolorbox}

\subsection{Analysis of a Real-World Incident}
\label{sec:rw_incident}

To demonstrate that a newly discovered invariant category in \name can act as a {practical oracle} to prevent a real-world exploit, we use Nomad Bridge incident (August 2022), a high-profile incident where a faulty upgrade broke a {proof-verification} check and enabled unauthorized message processing leading to asset theft (reported at $\approx\$190$M at the time of attack). The characteristics of this incident match with proof/signature verification (category~9) from RQ3 results. We use this invariant category to detect the vulnerability of Nomad incident~\cite{GCloudPostMortem,immunefi_hack_2023}.
Cluster~9 in RQ3 includes both (i) \emph{direct} cryptographic checks such as \texttt{MerkleProof.verify(...)} and (ii) \emph{derived proof-state} checks where successful verification is recorded into contract state (e.g., \texttt{proven[leaf]} or a stored root), and later enforced by a guard at the privileged entrypoint.
Nomad follows the second form: \texttt{process(m)} checks a stored proof marker (a mapping value) 
, and the upgrade bug made the default mapping value (\texttt{0x0}) appear acceptable. We distill a security invariant from this:
\[
  \forall m.\ \textsc{Process}(m)\ \Rightarrow\ \textsc{Proved}(m)
\]
i.e., \emph{across upgrades/initialization, a message must not be processed unless it has been proved under a trusted root}.

Nomad's cross-chain message acceptance follows the pattern of (i) a message is {proved} by verifying inclusion under a trusted Merkle root, then
(ii) the message is {processed} to release/mint assets.
The exploit root cause was an {upgrade-time misconfiguration} that made the ``unproved'' state indistinguishable from the ``proved'' state. This resulted in a routine upgrade marked a \code{bytes32(0)} root as trusted/committed, and because Solidity mappings default to zero values, messages that were never proved could appear as if they were proved under the zero root.
This enabled the attacker to call the privileged function \code{process(message)} without a valid proof step and still succeed, allowing for the mass draining of assets.

\textbf{Building a Foundry fuzzer.}
We implement an {upgrade-aware} fuzzer using Foundry's {stateful invariant testing}. 
We wrap the Nomad's message-processing contract in a Foundry {handler} that exposes three actions:
\textsc{Upgrade}, \textsc{Prove}, and \textsc{Process}.
During an invariant-based fuzzing campaign, Foundry repeatedly generates {randomized sequences} of calls to these actions, and checks the invariant after the sequence steps.
In our configuration, we do not add any Nomad-specific guidance. 
Instead, \textsc{Upgrade} takes its parameters directly as fuzz inputs (a \texttt{bytes32} root and a \texttt{uint64} timestamp-like value), while \textsc{Process} takes an arbitrary fuzzed byte-string as the message.
To make this baseline practical, we rely only on Foundry's {standard dictionary bias} for fuzzing, which biases generated inputs toward interesting boundary values such as \texttt{0x0} and small integers.
The fuzzer repeatedly generates {random call sequences} to these actions; for each call it also generates {random arguments} and a {random sender}. After each step (or at the end of the sequence, depending on configuration), Foundry evaluates the invariants and shrinks failing sequences.
Cluster~9 translates to a fuzzing oracle where a message must not be processed unless it has been proven under a trusted root/signature.
We enforce this oracle by instrumenting the handler's \textsc{Process} action. Before calling \code{process(message)}, we compute the message hash and query the contract's proof state. 
If the message is ``{unproved}'' (mapping default \code{0x0}), but \code{process} nevertheless succeeds, we set a ghost flag \code{unprovenProcessSucceeded = true}.
The invariant \code{invariant\_unprovenMessagesNeverProcess\_noHints()} asserts that this flag never becomes true.
This converts \name's proof-verification category into a concrete fuzzing oracle.

\textbf{Running the fuzzer.}
After running our fuzzer, it identifies a violation of invariants, reports the failing call sequence, and applies its shrinking procedure to minimize it, obtaining the smallest call sequence corresponding to this counterexample.
In our experiment, shrinking consistently yields a {two-step} counterexample that mirrors the real incident's causal structure:
\[
  \textsc{Upgrade}(\texttt{root}=0x0,\ \texttt{confirmAt}=1)\ ;\ \textsc{Process}(m)
\]
where $m$ is a fuzz-generated message that was never proved.
Foundry's trace output provides evidence for the oracle violation: immediately before \code{process}, the contract reports the message's proof state as \code{0x0} (unproved), yet \code{process} returns successfully instead of reverting.

\refstepcounter{finding}
\begin{tcolorbox}[title=RQ4 Answer: Effectiveness in Vulnerability Detection, flamesfinding, label=finding:meaningfulness]
Our case study demonstrates \name's newly discovered \emph{proof/signature verification} category can be leveraged as a concrete, reusable security oracle for  
a fuzzer to detect the logical vulnerability of the Nomad Bridge incident.
\end{tcolorbox}

\section{Discussion}
\label{sec:discussion}

This section discusses the results, their implications, and addresses limitations and threats to validity.


\subsection{Impact of Fine-Tuned Embeddings on Clustering}

\rb outperforms both general-purpose 
CodeBERT and domain-specific SmartBERT-v2 via invariant-based fine-tuning beyond contract-level understanding. This is consistent with the literature's focus on domain-specific fine-tuned BERT models for security~\cite{SecureBERT}. Furthermore, this shows task-specific fine-tuning is effective even with small but high-quality datasets (\num{1932} unique invariants), suggesting invariant analysis should prioritize specialized training over sole reliance on pre-trained models~\cite{LIMA}.

\subsection{The Predicate-Only Ablation}

Predicate-only views outperform predicate+message, which indicates error messages introduce lexical noise via common message templates,
creating spurious similarity across semantically distinct checks while separating semantically identical checks with different messages. Messages increase coverage 6\% (52\% $\rightarrow$ 58\%) by capturing edge cases but degrade cluster homogeneity. For taxonomy construction, which requires semantic coherence, predicate-only views are preferable.

\subsection{Reasons for the Coverage-Quality Trade-off}
The stark difference between DBSCAN's conservative assignment (52-58\% coverage, Silhouette 0.93) and HDBSCAN's inclusive approach (72-86\% coverage, Silhouette 0.73-0.85) reflects the precision-recall tension in cluster assignment. {DBSCAN's noise labeling acts as a semantic confidence threshold where border points between clusters remain unassigned rather than unreliably clustered}.
This conservativeness is appropriate for \name's goal. When claiming a cluster represents, e.g., ``replay prevention checks,'' we need high confidence that the majority of members share that intent. Manual analysis reveals that the \num{48}\% classified as noise also show patterns of (1) micro clusters or (2) multi-clause invariants detailed below.

\noindent\textbf{Micro clusters.}
Manual review reveals multiple micro-clusters within the unclustered invariants of our best configuration. For instance, \code{tx.gasprice > amount}, \code{tx.gasprice > _gas}, \code{tx.gasprice > nefwhitelist}, and \code{tx.gasprice > _rfee[from] && _rfee[from] != 0} could be clustered together, but our best configuration's conservativeness leaves them as noise. Review of the results for the configuration (SmartBERT embedding, HDBSCAN, predicate-only view) reveals these four invariants are misplaced into another cluster\footnote{\url{https://github.com/mojtaba-eshghie/Raven/blob/main/clustering/experiments/without/hdbscan_SmartBert.txt}}
, and for the other high-performing \name configuration, which uses predicate+messages view, one of these invariants is mis-clustered and the others are not clustered. Only K-Means configurations successfully generate micro-clusters. However, K-Means generates too many clusters and breaks down semantically relevant clusters into multiple groups making it unsuitable. Furthermore, some invariants that conduct very specialized checks, such as \code{uniswapv2pair != from && uniswapv2pair != to}, are under-represented in the dataset, and are semantically different from any other cluster; thus, they remain unclustered. The same holds for protocol-specific business logic invariants such as \code{key == uint256(uint160(msg.sender)) * (epoch()**2) + 1}. 


\noindent\textbf{Multi-clause invariants.}
We investigated whether the logical complexity of invariants affects clusterability by examining multi-clause invariants—those with multiple sub-predicates combined via Boolean operators (\code{&&}, \code{||}) or multiple comparison operators (\code{>=}, \code{<}, \code{!=}). Such invariants encode multiple independent intents, potentially producing less coherent embeddings. However, multi-clause forms appear equally in both groups: \num{53} among unclustered invariants and \num{76} among clustered ones (\num{17.43}\% vs \num{17.08}\%). Hence, this structural property does not correlate with the low coverage of our best-performing configuration.


\subsection{Why Six Categories Were Previously Unnoticed}

We attribute the discovery of the six new categories to two contributing factors. (1) Template-driven blindness: Prior tools use predefined invariant templates and hand-crafted categories of invariants 
that excel at finding anticipated patterns but cannot discover out-of-distribution defenses. Our data-driven approach makes no {a priori} assumptions, discovering structure from observed behavior. (2) Exploit-centric bias: security research emphasizes offensive analysis~\cite{SoK,zhou_sok_2023}, biasing templates toward vulnerabilities in exploited contracts; while prevented attacks leave only silent reverted transactions. 

\subsection{Open Questions}
\label{sec:future_work}

Our work opens several research directions:
\emph{(1) Scaling to Full Transaction History.} The 20k sample analyzed here represents 0.3\% of the 6.31M failures in our 9-month window. Scaling \name\ to the full corpus—and extending back to Ethereum's genesis block—would reveal: (a) temporal evolution of defenses (when did slippage guards emerge?), (b) protocol-specific patterns (do DeFi protocols share defensive characteristics?), and (c) what rare but important checks exist in the noise?
\emph{(2) Cross-Chain Comparative Analysis.} Applying \name\ to other blockchain networks (e.g., Solana) would test whether defensive patterns are EVM-universal or chain-specific. Solana's account model and Rust-based constraints may produce different invariant families. Such a comparison can inform cross-chain security practices. \emph{ (3) Hybrid Neuro-Symbolic Approaches.} While \name\ uses neural embeddings for scalability, pairwise symbolic comparison (e.g., via \sindi~\cite{SindiGithub}) could validate intra-cluster semantic consistency. A hybrid approach that combines neural clustering for discovery and symbolic validation of the discovered invariant. This could reduce false positives in clustering membership.




\subsection{Threats to Validity}
\label{sec:limitations}

\subsubsection{Extraction Pipeline Accuracy} Our invariant extraction relies on Tenderly's dynamic analysis API, which has a \num{3}\% failure rate and a \num{13.8}\% source code mismatch with Etherscan (when the tool does not yield an invariant but the source code is present on Etherscan). Nevertheless, manual validation of \num{1000} transactions confirmed \num{100}\% accuracy for the {successfully extracted invariants}.

\subsubsection{Bias Towards Verified Contracts} The \num{23.3}\% ``No source code found'' exclusion creates survivor bias toward professionally developed and verified contracts.
While there is no definitive answer for the proportion of unverified contracts being malware, unverified source code is typically used as a red flag for malware detection~\cite{defiRugPulls}. 


\section{Related Works}
\label{sec:related_works}
The conceptual framework of invariants was established by 
Floyd~\cite{floydAssigning}. He introduced a method for reasoning about a program's correctness by associating assertions, known as invariants, with specific points in a control flow graph, such as condition branches or loops. This foundation was extended by Hoare Logic~\cite{Hoare}, which laid the groundwork for formal program verification by treating such assertions as conditions that must remain true throughout execution.
Related research on invariants and semantic clustering falls into three threads: dynamic mining and static synthesis of invariants, code representations, and empirical studies on blockchain's failed transactions. We position
\name against each of these.

\noindent\textbf{Smart contract invariant mining and synthesis.}
Trace2Inv~\cite{DemystifyingInvariants} infers invariants from transaction traces using templates and reports coarse categories (e.g., access control, data flow, time lock). InvCon~\cite{InvCon} and its successor, InvCon+~\cite{InvCon+}, target token-centric invariants (ERC-20 and ERC-721) via specification templates (i.e., catalogs of Solidity invariants) and transaction history. SmartOracle~\cite{SmartOracle} also mines pattern-based properties from transaction history. Eshghie et al.~\cite{captureDCR,FormalizingDCRMojtaba} offer a formalization of \num{15} business logic properties that encode many well-known smart contract invariants and are used for tasks such as monitoring and PoC exploit synthesis~\cite{highguard,xplogen}. InvSol~\cite{InvSol} transcompiles the Solidity contract to Java + JML, uses a test crafter to generate many test cases via which it infers Daikon-style invariants.
All mentioned works use invariant templates that do not benefit from the deployed successful defenses on Ethereum. Instead, they rely on hand-crafted template invariants that are not comprehensive and may not reflect the real-world defenses. 
\flames~\cite{flames} generates invariants in the form of pre-/post-conditions to harden contracts without vulnerability labels.
VeriSmart~\cite{VeriSmart} statically synthesizes logical invariants with SMT to prove or refute safety properties of smart contracts. Both \flames~\cite{flames} and VeriSmart~\cite{VeriSmart} produce invariants on \emph{source} or \emph{IR}, and do not characterize which defensive checks actually block real-world attacks. 

\noindent\textbf{Neural and symbolic code representations.}
General-purpose encoders, such as CodeBERT~\cite{CodeBERT}, provide contextual embeddings for code, whereas domain-specific models like SmartBERT-v2~\cite{smartbert2024} are tailored to Solidity sources. SolBERT~\cite{SolBERT} uses contrastive learning with AST views, and SmartEmbed~\cite{SmartEmbed} fuses CFG features to detect clone-related bugs. \sindi~\cite{SindiGithub} compares the semantic strength of pairs of invariants symbolically. These enable semantic embedding but either operate at contract granularity, rely on templates, or address {pairwise} relations rather than {global} structure. \name differs by fine-tuning an \emph{invariant-level} encoder (Solidity syntax) on failed transaction invariants and applying semantic clustering to reveal semantic invariant groups without labels.

\noindent\textbf{Empirical studies of failed transactions.}
Large-scale studies report why transactions fail and where gas is spent (e.g., out-of-gas, bot-driven spam)~\cite{CharacterizingTXRevertingSC,Solana}. These studies quantify failure modes but typically record error strings or high-level reasons, and do not extract and aggregate the precise invariants responsible for the reverts, nor do they build an invariant taxonomy. \name closes this gap by extracting invariant predicates responsible for the transaction failures, and semantically clustering them. \name enables a future large-scale study of all failed transactions (e.g., on Ethereum) to determine the most failure-inducing and potentially attack-reverting invariants effective in severe recent DeFi incidents~\cite{SoK}.

\section{Conclusion}\label{sec:conclusion}
This work presents \name, a tool to analyze and semantically cluster Ethereum's on-chain failed (reverted) transactions. 
We reframe reverted executions on Ethereum as a signal for understanding what defenses implemented by the contracts actually work on the chain. More specifically, the transactions that were reverted by \code{require/assert/if...revert/throw} statements (invariants) in the smart contract on Ethereum are the input to \name.
\name\ extracts the precise predicates that trigger reverts, embeds them with a contrastively fine-tuned BERT-based model, and semantically clusters them. On a representative sample of \num{20}k failed transaction failures (mapping to \num{727} unique invariants that revert them), \name yields well-separated clusters. 
A manual expert analysis of the generated \num{19} clusters confirms coherent defensive invariants aligned with prior taxonomies and identifies \emph{six} \emph{previously unreported} categories (e.g., replay prevention, signature/Merkle verification, allow/ban/bot lists, etc.), indicating semantic analysis reverted executions on blockchain reveals defense patterns missed by the literature.
Our case study on a real-world incident further validates the utility of these invariants as security analysis oracles.
We envision \name\ as a tool for researchers and security practitioners to map the ecosystem's \emph{working} defensive surface, informing invariant catalogs of invariant synthesizers/miners and analysis tools.

\bibliographystyle{IEEEtran}
\bibliography{refs}

\end{document}